\newcommand{\cs}[3]{{{#3} \brace {#1 #2}}}

\newcommand{\h}[1]{\mathop{\lambda}\limits_{#1}\ \!\!\!}

\newcommand{\edf}{\ {\mathop{=}\limits^{\rm def}}\ }

\documentclass{article}
\begin{document}

\begin{center}
\textbf{Dark Energy: Is It of Torsion Origin? }

\it{M.I.Wanas$^{1,2}$}

$^1$Astronomy Department, Faculty of Science, Cairo
University, Giza, Egypt.\\
E-mail:wanas@mailer.eun.eg\\
$^2$ The Egyptian Relativity Group (ERG),\\ {\bf URL}:erg.eg.net
\end{center}

\begin{abstract}
{\it "Dark Energy"} is a term recently used to interpret
supernovae type Ia
 observation. In the present work we give two arguments on a
 possible  relation between dark energy and torsion of space-time.
\end{abstract}
\section {Introduction}

Recently, the exotic term {\it "Dark Energy"} is frequently used
to interpret supernovae type Ia observations [1], [2]. These
observation indicate very clearly that the Universe is in an
accelerating expansion phase. This implies the existence of a
large scale {\bf "repulsive force"}, causing the observed
accelerating expansion phenomena.

It is well known that the interaction responsible of the behavior of
the large scale structure and evolution of the Universe is mainly
gravitational. Unfortunately, gravity as we understand it in the
solar, and comparable, systems cannot account for supernovae type Ia
observations. The reason is that gravity theories, including general
relativity (GR), deals with gravity as an {\bf "attractive force"}.

In the literature, it is widely accepted that inserting a cosmological
term in Einstein's field equations can solve the problem of supernovae
type Ia observation. But the cosmological constant itself has many problems,
 concerning its value which is still controversial [3]. So, it is necessary
 to seek other interpretations, to solve this problem.

Einstein, in constructing his theory of GR, has used a geometric
property, {\it "the curvature"}, to interpret gravity. Several
applications and predictions of this theory show its success over
about eight decades. I consider this success as a success of using
geometry in solving physical problems, rather than a success of GR
itself. So, when new problems concerning interpretation of large
scale phenomena appear, I prefer to revisit geometry, seeking a
solution. Any geometric structure, characterized by a linear
connection, has two important geometric entities: {\it "Curvature"
and "Torsion"}. Einstein has used the curvature to interpret,
successfully, attractive gravity. What is the role of torsion?

The aim of the present work is to review briefly some of the
properties of the torsion, and to show that it gives rise to a
repulsive force, in contrast to the attractive force implied by the
curvature. For this reason in section 2, I briefly review a
geometric structure with simultaneously non-vanishing curvature and
torsion. In section 3, I give some properties of the
torsion giving rise to a repulsive force. The paper is discussed and
concluded in section 4.

 \section{Geometries with  Curvature and Torsion: The PAP-Geometry}
Torsion tensor is the antisymmetric part of any non-symmetric
linear connection. Geometries with curvature and torsion are
frequently used (see references [4], [5]) and are classified in
the literature as {\it "Riemann-Cartan"} geometry. In what follows, we are
going to review very briefly the main features of a version of
Absolute Parallelism (AP)-geometry, {\it "The Parameterize
Absolute Parallelism (PAP)-Geometry"}, in which both torsion and
curvature are simultaneously non-vanishing. We have chosen this
type of geometry since calculations in its context are very easy (
for details, see reference [6]).

The structure of a 4-dimensional PAP-space is defined completely
by a set of 4-contravariant linearly independent vector fields
$\h{i}^{\mu}$. This geometry is characterized by the linear
non-symmetric connection,
$$
\nabla^{\alpha}_{. \mu \nu} = \cs{\mu}{\nu}{\alpha} + b ~
\gamma^{\alpha}_{. \mu \nu}, \eqno{(3)}
$$
where $\cs{\mu}{\nu}{\alpha}$ is Christoffel symbol of the second
kjnd, $\gamma^{\alpha}_{. \mu \nu}$ is the contorsion and $b$ is a
dimensionless parameter. The connection (3) is metric and has a
non-vanishing torsion ($\hat{\Lambda}^{\alpha}_{. \mu \nu} = b
\Lambda^{\alpha}_{. \mu \nu} \edf \nabla^{\alpha}_{. \mu \nu} -
\nabla^{\alpha}_{. \nu \mu}$). The curvature tensor corresponding
to (3) is ,in general, non-vanishing and defined by
$$
{\hat{B}}^{\alpha}_{.\mu \nu \sigma}= R^{\alpha}_{.\mu \nu
\sigma}+ b~ \hat{Q}^{\alpha}_{.\mu \nu \sigma},   \eqno{(4)}
$$
where,
$$
R^{\alpha}_{~\mu \nu \sigma} \edf \cs{\mu}{\sigma}{\alpha}_{, \nu} -
\cs{\mu}{\nu}{\alpha}_{, \sigma}+
\cs{\mu}{\sigma}{\epsilon}\cs{\epsilon}{\nu}{\alpha}
-\cs{\mu}{\nu}{\epsilon}\cs{\epsilon}{\sigma}{\alpha}, \eqno{(5)}
$$
is  the Riemann- Christoffel curvature tensor and the tensor
$\hat{Q}^{\alpha}_{\mu \nu \sigma}$ is defined by
$$
\hat{Q}^{\alpha}_{~\mu \nu \sigma} \edf
\gamma^{\stackrel{\alpha}{+}}_{~
\stackrel{\mu}{+}\stackrel{\sigma}{+}|\nu}-\gamma^{\stackrel{\alpha}{+}}_{~
\stackrel{\mu}{+}\stackrel{\nu}{-}|\sigma}- b(\gamma^{\beta}_{~
\mu \sigma}\gamma^{\alpha}_{\beta \nu} -  \gamma^{\beta}_{~ \mu
\nu}\gamma^{\alpha}_{\beta \sigma}). \eqno{(6)}
$$
The stroke and the (+)-sign are used to characterize covariant
derivatives using the linear connection (3), while the stroke and
the (-)-sign are used to characterize covariant derivatives using
the dual connection $\tilde{\nabla}^{\alpha}_{. \mu \nu} (=
\nabla^{\alpha}_{. \nu \mu})$.

The path equations corresponding to (3) can be written as,
$$
\frac{d^{2}x^{\mu}}{d\tau^{2}} +
\cs{\alpha}{\beta}{\mu}\frac{dx^{\alpha}}{d\tau}
\frac{dx^{\beta}}{d\tau} = - b~~\Lambda^{..\mu}_{\alpha
\beta.}\frac{dx^{\alpha}}{d\tau} \frac{dx^{\beta}}{d\tau},
\eqno{(7)}
$$
where $\tau$ is a scalar parameter.

It is to be noted that the PAP-geometry covers both Riemannian
geometry ($b=0$) and conventional AP-geometry ($b=1$), as special
cases.

\section{Dark Energy-Torsion Relation }
In this section we give two arguments on a probable relation between
torsion and dark energy. The first is a kinematical argument and the second is a dynamical one.\\ \\
{\bf The Kinematical  Argument:} In the case $b=1$ (the AP-case [6]),
the case of vanishing curvature, equation (4) gives,
$$
 R^{\alpha}_{.\mu \nu \sigma}
\equiv - Q^{\alpha}_{.\mu \nu \sigma}.   \eqno{(8)}
$$
Although the tensors $R^{\alpha}_{. \mu \nu \sigma}$ and
$Q^{\alpha}_{. \mu \sigma \nu }$ appear to be mathematically
equivalent, they have the following differences:\\
1- The Riemann-Christoffel curvature tensor is made purely from
Christoffel symbols (see (5)), while the tensor $Q^{\alpha}_{.\mu
\nu \sigma}$ (given by (6) with $ b = 1$) is made purely from the
contortion or from the torsion via the relation
$$
\gamma^\alpha_{. \mu \nu} = \frac{1}{2}(\Lambda^\alpha_{. \mu \nu}
-\Lambda^{.\alpha.}_{\mu . \nu} - \Lambda^{.\alpha.}_{\nu .\mu })
$$ The first
tensor is non-vanishing in Riemannian
geometry, while the second vanishes in the same geometry. \\
2- The non-vanishing of $R^{\alpha}_{.\mu \nu \sigma}$ is the
measure of the curvature of the space, while the addition of
$Q^{\alpha}_{.\mu \nu \sigma}$ to it, causes the space to be flat.
So, one is causing an inverse effect, on the properties of
space-time, compared to the other. For this reason we call
$Q^{\alpha}_{.\mu \nu \sigma}$ (or $\hat{Q}^{\alpha}_{.\mu \nu
\sigma}$)"{\underline{\it{The Curvature Inverse of
Riemann-Christoffel Tensor}"}} [6], or in other
words\\{\underline{\it "The Additive inverse of the Curvature
Tensor"}}. Note that both tensors are considered as curvature
tensor, but one of them cancels the effect of the other, if both
existed in the same geometric structure.

Now, in view of the above two differences, we can deduce that these
two tensors are not, in general, equivalent. In other words, if we
consider gravity as curvature of space-time and is represented by
$R^{\alpha}_{. \mu \nu \sigma}$, we can consider $Q^{\alpha}_{.\mu
\nu \sigma}$ as representing {\bf{anti-gravity}}! The existence of
equal effects of gravity and anti-gravity in the same system
{\it neutralizes
} the space-time (make it flat), geometrically. This
situation is similar to the existence of equal quantities of
positive and negative electric charges in the same system, which
neutralizes the system, electrically.

Let us now approach, geometrically, the problem of dark energy. It
is well known that Riemann-Christofel curvature tensor (5)
satisfies Bianchi second identity, which can be written in the
contracted form
$$
(R^\mu_{. \nu} - \frac{1}{2} \delta^\mu_{. \nu} R); \mu = 0.
\eqno{(9)}
$$
This identity is interpreted, physically as a generalization of a
law of conservation. The quantity between brackets in (9)
represents the conserved quantity. We can attribute to this
quantity the property of energy associated  with  curvature, {\it
the "Curvature Energy"}. Similarly, using (8) \& (9), we can
write,
$$
(Q^\mu_ {. \nu} - \frac{1}{2} \delta^\mu_{. \nu} Q); \mu = 0.
\eqno{(10)}
$$
where $Q^\mu_ {. \nu}$ is the only non-vanishing contracted  form
of (6) (with $b = 1$ ) and $Q$ is its scaler. Now, using the same
argument, (10) can be considered as conservation of another type
of energy represented by the quantity between brackets, and since
$Q^\mu_ {. \nu}$ and its contraction $Q$ are made purely from the
contortion (or the torsion), as clear from (6), we call this type
of energy the {\it "Torsion Energy"}.

If we assume that the effects of gravity and anti-gravity are not
exactly equal in the same system, then space-time curvature can be
represented by the tensor (4). The existence of anti-gravity gives
rise to a repulsive force, which can be used to interpret SN type
Ia observation. This can be achieved by adjusting the parameter
$b$ .

 It is clear that,  "Torsion Energy" follows a  conservation
law, similar to that of the curvature energy (for details see
reference [7])\\ \\
{\bf The Dynamical Argument} for the existence of a repulsive
force, corresponding to the torsion of space-time, consider the
linearized form of (7) which can be written as [8],
$$
\Phi_{T} = \Phi_{N} (1-b)= \Phi_{N} + \Phi_{\Sigma}, \eqno{(11)}
$$
where,
$$
\Phi_{\Sigma} \edf - b \Phi_{N}, \eqno{(12)}
$$
 $\Phi_{N}$ is the Newtonian gravitational potential and
$\Phi_{T}$ is the total gravitational potential due the presence of
gravity and anti-gravity. It is clear from (11) that the Newtonian
potential is reduced by a factor $b$ due to the existence of the
torsion energy. It is obvious from (12) that $ \Phi_{\Sigma} $ and
the Newtonian potential have opposite signs. Then one can deduce
that $\Phi_{\Sigma}$ is a {\it repulsive potential}, in contrast to
the {\it attractive potential} $\Phi_{N}$.
\section{Discussion and Concluding Remarks}
In the present work we have chosen a version of the 4-dimensional
AP-geometry to represent the physical world including space and
time. On the present work, we can draw the following remarks:\\
1- It is clear that {\it torsion energy}, defined in the previous
section, can solve the problem of SN type Ia observations, since it
gives rise to a repulsive force. This can be achieved by adjusting
the parameter $b$. One can now replace the exotic term {\it{"dark
energy"}} by the term {\it{"torsion energy"}}. The later has a pure
geometric origin.\\
2- Curvature and torsion corresponds to two different types of
energy. The energy corresponding to the first gives rise to an
attractive force, while the energy corresponding to the second is
repulsive. We believe that torsion energy is what has been
discovered recently by the SN type Ia observation [1]. Both energies
obey the same conservation.\\
3- The results obtained in the present work can be obtained, with
some efforts, using other geometries with curvature and torsion,
since such geometries possess similar features [9].\\
4- From the geometerization point of view, one can conclude that
"Dark Energy" is nothing but "Torsion Energy", responsible for
repulsion in the Universe. For more details and discussions cf.
[10], [11], [12]\& [13].
\section*{References}
\noindent
1. J.L. Tonry, B.P. Schmdit, et al. {\it Astrophys. J.}{\bf{594}}, 1 (2003). \\
2. P.D. Mannheim, {\it Prog.Part.Nucl.Phys.} {\bf{56}} 340 (2006); gr-qc/0505266 \\
3. S.M. Carroll, {\it"The Cosmological Constant"},
http://livingreviews.org/
Irr-2001-1. \\
4. A. Einstein, {\it "The Meaning of Relativity"}
(Princeton, 5th ed. 1955).\\
5. A. Einstein, {\it Math. Annal.} {\bf 102}, 685 (1930).\\
6. M.I.Wanas, {\it Turk. J. Phys.} {\bf{24}}, 473 (2000); gr-qc/0010099 \\
7. M.I. Wanas, arXiv:0705.2255 (2007). \\
8. M.I. Wanas, {\it Astrophys. Space Sci.}, {\bf{258}}, 237 (1998); gr-qc/9904019 \\
9. M.I. Wanas and M.E. Kahil, {\it Gen.Rel.Gravit.} {\bf 31} 1921 (1999); gr-qc/9912007. \\
10. M.I. Wanas, {\it Proc. MG XI}, Part{\bf{B}}, 1782 (2008); arXiv:0704.3760\\
11. M.I. Wanas, {\it Int. J. Mod. Phys. A},{\bf 22},5709-5716 (2007); arXiv:0802.4104 \\
12. M.I. Wanas, {\it Proc. NUPPAC'7}, Page 1, (2008); arXiv:0809.5040 \\
13. M.I. Wanas, Proc. {\it DSU IV}, AIP Conf. series {\bf 1115}, 218 (2009). \\
\end{document}